\nonstopmode

\documentclass[prl,twocolumn]{revtex4}
\usepackage{graphicx}
\usepackage{natbib}
\usepackage{epstopdf}
\usepackage[hidelinks, colorlinks = true, urlcolor=blue, linkcolor=blue, citecolor= blue]{hyperref}
\usepackage{color}
\usepackage{amsmath}
\usepackage{lettrine}
\usepackage{lipsum}
\begin{document}

\title{Buckling-Induced Kirigami}

\author{Ahmad Rafsanjani$^1$ and Katia Bertoldi$^{1,2}$}
\email[Correspondence to ]{bertoldi@seas.harvard.edu}
\affiliation{$^1$John A. Paulson School of Engineering and Applied Sciences, Harvard University, Cambridge, MA 02138, USA}
\affiliation{$^2$Kavli Institute, Harvard University, Cambridge, MA 02138, USA}

\date{\today}

\hyphenpenalty=5000

\begin{abstract}

We investigate the mechanical response of thin sheets perforated with a square array of mutually orthogonal cuts, which leaves a network of squares connected by small ligaments. Our combined analytical, experimental and numerical results indicate that under uniaxial tension the ligaments buckle out-of-plane, inducing the formation of 3D patterns whose morphology is controlled by the load direction. We also find that by largely stretching the buckled perforated sheets, plastic strains develop in the ligaments. This gives rise to the formation of kirigami sheets comprising periodic distribution of cuts and permanent folds. As such, the proposed buckling-induced pop-up strategy points to a simple route for manufacturing complex morphable structures out of flat perforated sheets.
\end{abstract}

\maketitle


In recent years, origami~\cite{Mahadevan2005, Wei2013, Schenk2013, Tachi2013, Silverberg2014, Yasuda2015, Silverberg2015, Overvelde2016, Dudte2016} and kirigami~\cite{Castle2014, Isobe2016,Sussman2015, Eidini2015, Chen2016, Tang2016, Seffen2016, Shyu2015, Blees2015, Song2015, Lamoureux2015, Wu2016, Zhang2015, Eidini2016, Castle2016, Neville2016, Yan2016, Norman1993} have become emergent tools to design programmable and reconfigurable mechanical metamaterials.
Origami-inspired metamaterials are created by folding thin sheets along predefined creases, whereas kirigami allows the practitioner to exploit cuts in addition to folds to achieve large deformations and create 3D objects from a flat sheet.
Therefore, kirigami principles have been exploited to design highly stretchable devices~\cite{Norman1993,Song2015,Shyu2015,Blees2015,Lamoureux2015,Wu2016, Isobe2016} and morphable structures~\cite{Zhang2015, Yan2016, Neville2016}.
Interestingly, several of these studies also show that pre-creased folds are not necessary to form complex 3D patterns, as mechanical instabilities in flat sheets with an embedded array of cuts can result in out-of-plane deformation ~\cite{Norman1993,Shyu2015, Blees2015, Lamoureux2015, Isobe2016,Zhang2015, Wu2016,Yan2016}. However, while a wide range of 3D  architectures have been realized by triggering buckling under compressive stresses~\cite{Zhang2015, Yan2016}, instability-induced kirigami designs subjected to tensile loading are limited to a single incision pattern comprised of parallel cuts in a centered rectangular arrangement~\cite{Shyu2015, Blees2015, Lamoureux2015, Norman1993, Isobe2016}.

\begin{figure}[b]
\centering
\linespread{0.8}
\includegraphics [width=\columnwidth]{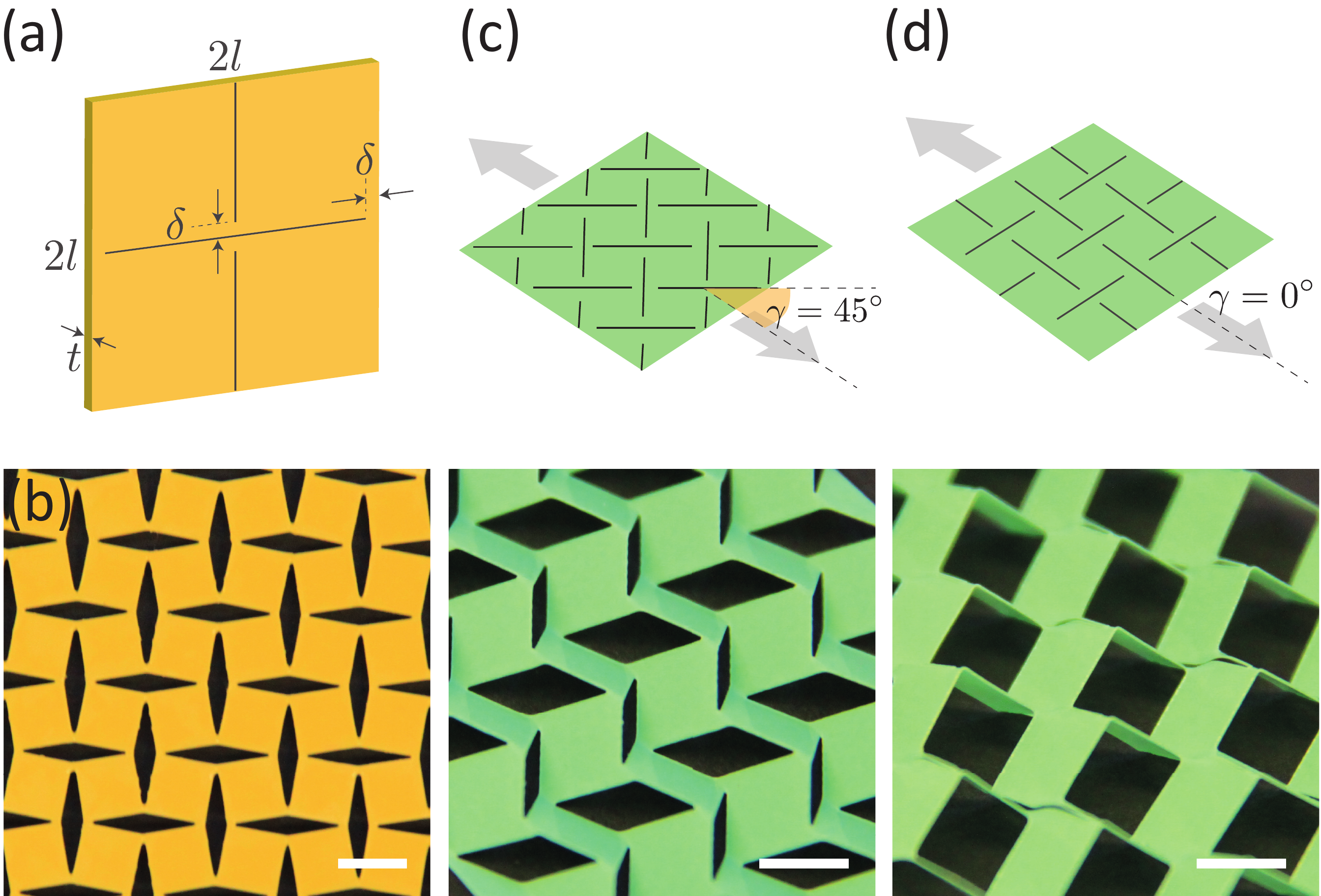}
 \caption{ (a) Schematic of the system: an elastic sheet of thickness $t$ perforated with a square array of mutually orthogonal cuts. (b) In the thick limit (i.e. for large values of $t/\delta$) the perforated sheet deforms in plane and identically to a network of rotating squares~\cite{Grima2000}. (c-d) For sufficiently small  values of $t/\delta$ mechanical instabilities triggered under uniaxial tension result in the formation of complex 3D patterns, which are  affected by the loading direction. The 3D patterns obtained for $\gamma=45^\circ$ and  $\gamma=0^\circ$ are shown in (c) and (d), respectively.
 Scale bars: 6~mm. }
\label{Fig1}
\end{figure}

\begin{figure*}[t]
\centering
\linespread{1}
\includegraphics [width=\textwidth]{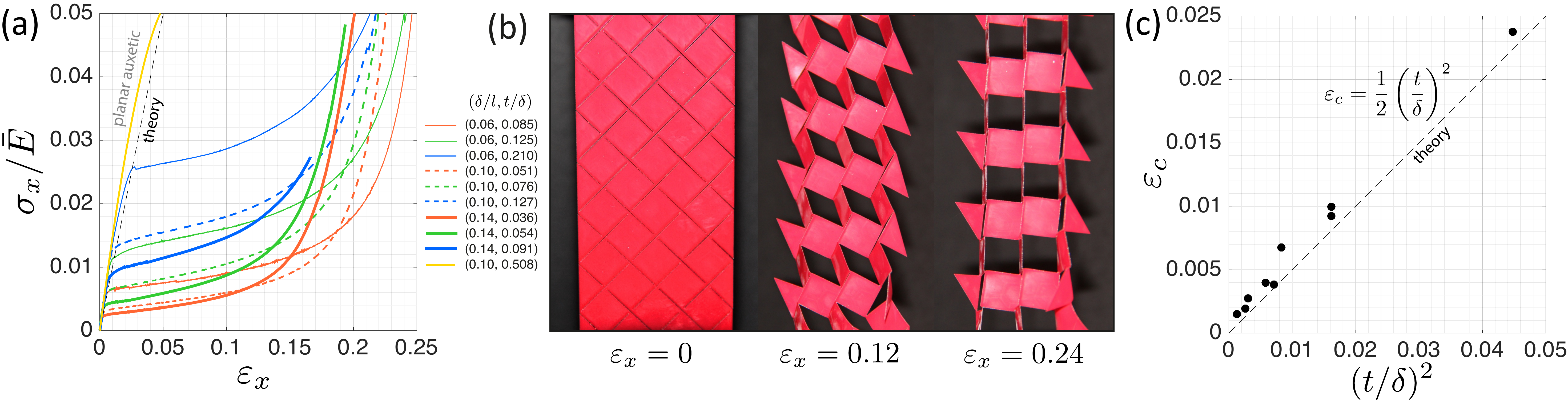}
 \caption{ (a) Experimental stress-strain curves  for perforated sheets characterized by different normalized hinge width $\delta/l$ and normalized sheet thickness $t/\delta$ for $\gamma=45^\circ$. Note that the stress is normalized by the effective in-plane Young's modulus $\bar{E}=2/3 E (\delta/l)^2$.
(b)  Snapshots of the sample  with  $\delta/l=0.06$ and $t/\delta\simeq0.085$ at $\varepsilon_x=0$, 0.12 and 0.24.
(c) Critical strain $\varepsilon_c$ as a function of $(t/\delta)^2$ as obtained from experiments (markers) and predicted analytically (dashed line).}
\label{Fig2}
\end{figure*}

In this Letter, we investigate the tensile response of elastic sheets of thickness $t$ perforated with a square array of mutually orthogonal cuts.
This perforation pattern introduces a network of square domains of edge $l$ separated by hinges of width $\delta$ [Fig.~\ref{Fig1}(a)]. While the planar response of such perforated sheets in the thick limit (i.e. for large values of $t/\delta$) has received significant attention, as it is characterized by effective negative Poisson's ratio~\cite{Grima2000,Grima2005,Grima2011,Cho2014, Gatt2015, Shan2015, Suzuki2016, Vasiliev2002, Rafsanjani2016} [Fig.~\ref{Fig1}(b)], here we add another dimension and study how the behavior of the system evolves when  the thickness is progressively decreased (i.e. for decreasing values of $t/\delta$). Our combined analytical, numerical and experimental results indicate that in sufficiently thin sheets mechanical instabilities triggered under uniaxial tension can be exploited to create complex 3D patterns and even to guide the formation of permanent folds. We also find that the morphology of the instability-induced patterns is strongly affected by the loading direction (see Figs.~\ref{Fig1}(c)-(d) and  Movies 1 in Supplemental Material~\cite{SI}), pointing to an effective strategy to realize  functional surfaces characterized by  a variety  of  architectures.


We start by experimentally investigating the effect of the sheet thickness $t$ and hinge width $\delta$  on the response of the system subjected to uniaxial tension along the square diagonals  [i.e. for $\gamma=45^\circ$ - Fig.~\ref{Fig1}(c)].
Specimens are fabricated by laser cutting an array of $3\times8$ mutually perpendicular cuts [see Fig.~\ref{Fig2}(b)] into plastic sheets (Artus Corporation, NJ) with Young's modulus $E=4.33\:\text{GPa}$ and Poisson's ratio $\nu\simeq0.4$ (see Supplemental Material: {\it Experiments}~\cite{SI}).
In Fig.~\ref{Fig2}(a), we report the experimental stress-strain responses for 10 samples characterized by different values of normalized thickness $t/\delta$ and normalized hinge width $\delta/l$.

First, it is apparent that the initial response for all samples is linear.
At this stage, all hinges bend in-plane, inducing  pronounced
rotations of the square domains [Fig.~\ref{Fig1}(b)], which  result in large negative values of the macroscopic Poisson's ratio~\cite{Vasiliev2002,Grima2005}.
As such, the stiffness of the perforated sheets, $\bar E$,  is governed by the in-plane flexural deformation of the hinges and it can be shown that (see Supplemental Material: {\it Analytical Exploration}~\cite{SI}):
\begin{equation}
\centering
\bar{E}=\frac{\sigma_x}{\varepsilon_x}=\frac{2}{3}E \left(\frac{\delta}{l} \right)^2.
\label{Eq1}
\end{equation}

Second, for the thin samples (i.e. $t/\delta\ll1$), the curves reported in Fig.~\ref{Fig2}(a) also show a sudden departure from linearity to a plateau stress caused by the out-of-plane buckling of the hinges. Such buckling in turn induces out-of plane rotations of both the square domains and the cuts, which arrange to form a 3D pattern reminiscent of a misaligned Miura-ori~\cite{Saito2016} with an alternation of square solid faces (corresponding to the square domains) and rhombic open ones (defined by the cuts) (see Fig.~\ref{Fig1}(c),  Fig.~\ref{Fig2}(b) at $\varepsilon_x=0.12$ and Movie 2 in Supplemental Material~\cite{SI}). To characterize the critical strain, $\varepsilon_c$, at which the instability is triggered, we start by noting that since the stress  immediately after instability is almost constant, the contribution of out-of-plane strain energy $\mathcal{U}_o$ should be linear in $\varepsilon_x$, (see Supplemental Material:
\emph{Analytical exploration}~\cite{SI})
\begin{equation}
\centering
\mathcal{U}_o(\varepsilon_x)=\bar{E}\varepsilon_c(\varepsilon_x-\varepsilon_c).
\label{Eq2}
\end{equation}
Moreover, assuming that the square domains remain rigid and that the deformation localizes at the hinges which can be modeled as flexural beam segments, $\mathcal{U}_o$ can also be written as
\begin{equation}
\centering
\mathcal{U}_o(\varepsilon_x)=8\times\frac{1}{8 l^2t}\int_0^\delta  \frac{E I_o}{\rho_o^2} ds=\frac{1}{3}E\left(\frac{t}{l} \right)^2\theta_o^2,
\label{Eq3}
\end{equation}
where $I_o=\delta t^3/12$, $\rho_{o}=\delta/2\theta_{o}$ and $2\theta_o$ is the opening angle of each cut after out-of-plane buckling, which for $\gamma=45^0$ is approximated by
\begin{equation}
\centering
\theta_o^2\simeq\varepsilon_x-\varepsilon_c.
\label{Eq4}
\end{equation}
Finally, by equating Eqs. (\ref{Eq2}) and (\ref{Eq3}) we find that
\begin{equation}
\centering
 \varepsilon_c \simeq\frac{1}{2} \left ( \frac{t}{\delta} \right )^2,
\label{Eq5}
\end{equation}
which despite the simplifications made, compares very well with our experimental results [Fig.~\ref{Fig2}(c)] and numerical simulations [Fig.~S6]. Note that a similar expression for the critical strain has been previously obtained for kirigami patterns comprising  parallel cuts in a centered rectangular arrangement~\cite{Isobe2016}.
\begin{figure}[ht]
\centering
\linespread{1}
\includegraphics [width=\columnwidth]{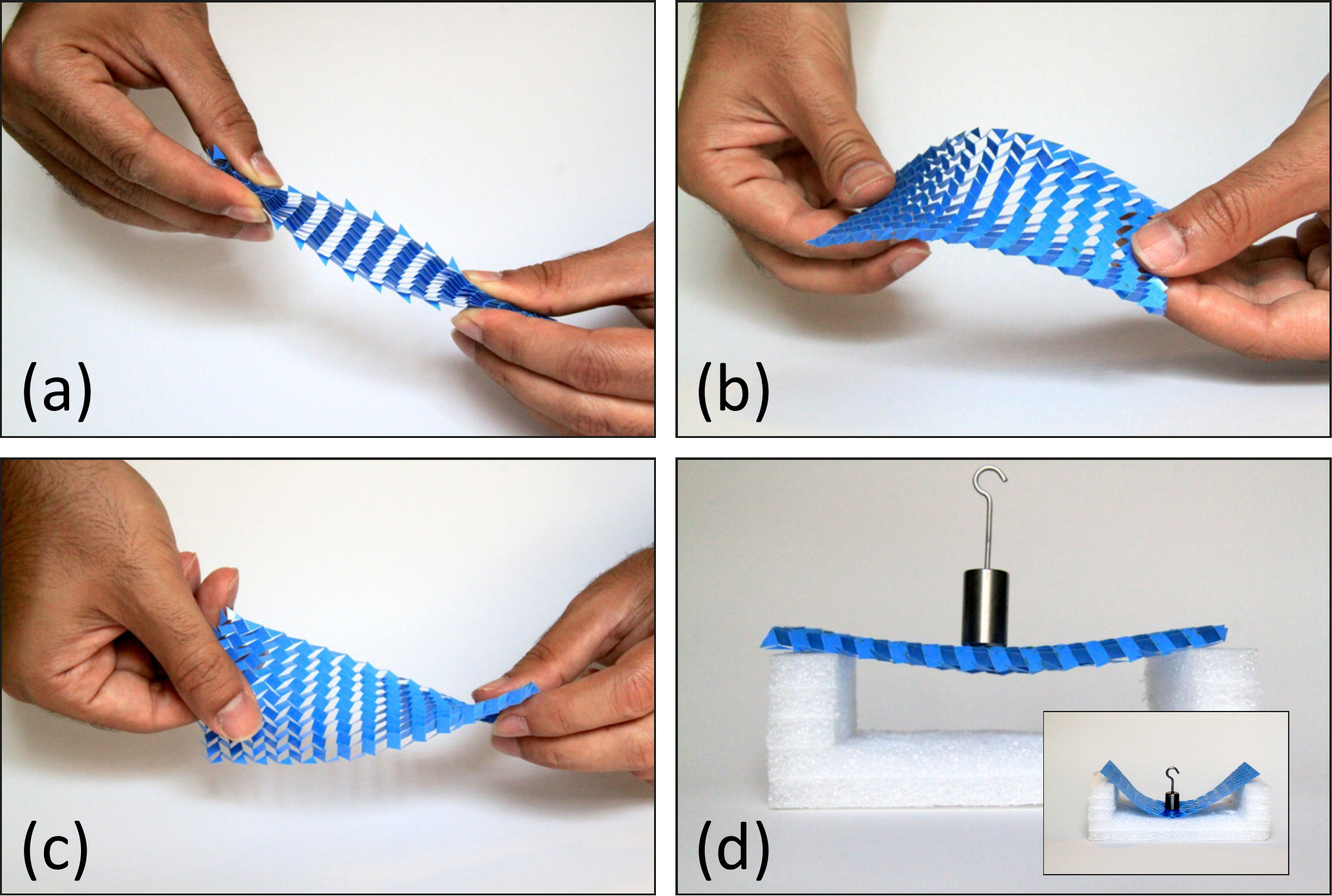}
 \caption{The buckling-induced Miura kirigami sheet (a) is flat-foldable, (b) forms a saddle shape with a negative Gaussian curvature upon non-planar bending, (c) twists under anti-symmetric out-of-plane deformation
(d) has much higher bending rigidity than the corresponding flat perforated sheet (inset).
Note that the 127~$\mu$m thick Miura kirigami sheet shown here supports a 20g weight. }
\label{Fig3}
\end{figure}

\begin{figure*}[t]
\centering
\linespread{1}
\includegraphics [width=0.9\textwidth]{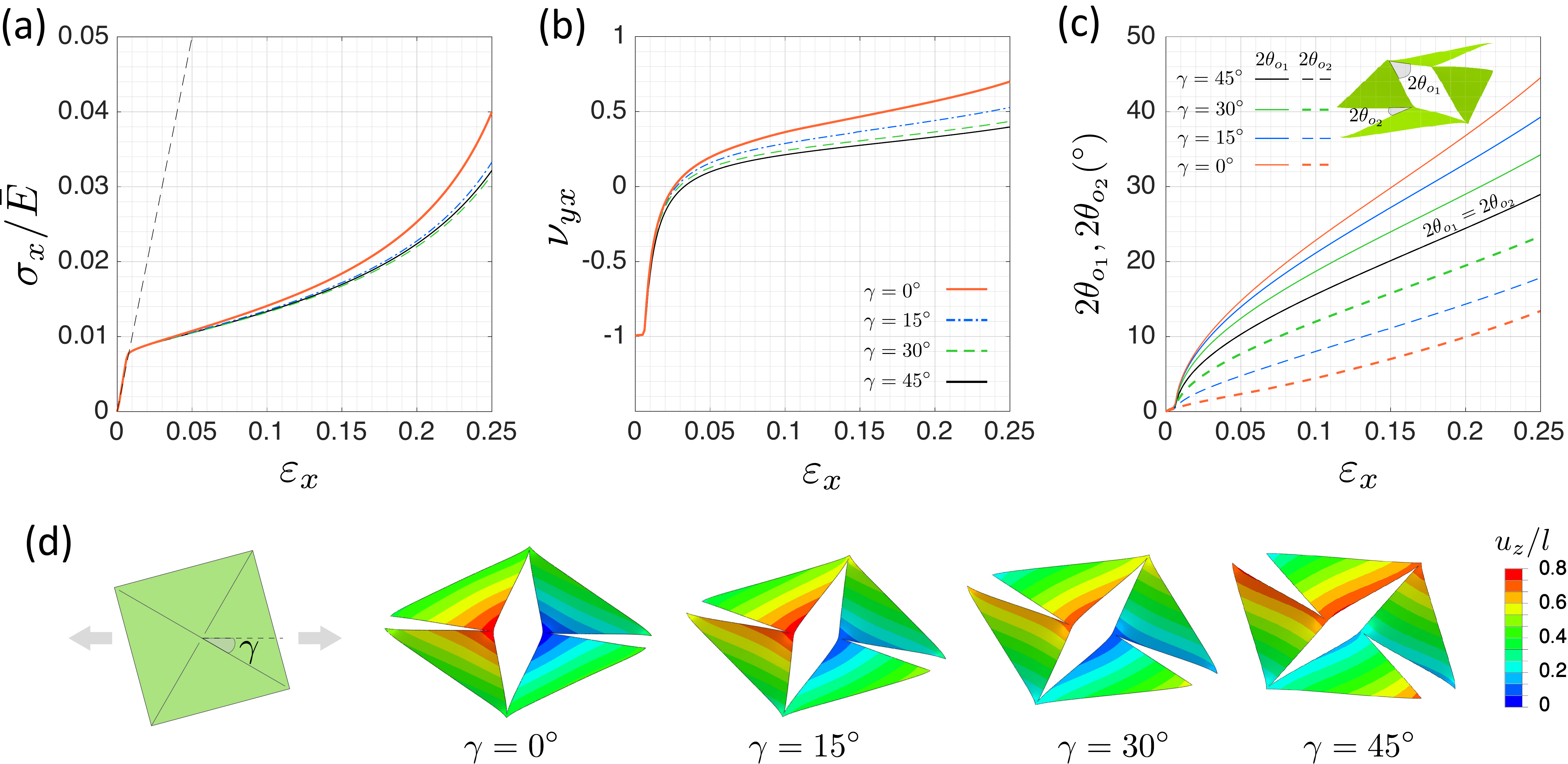}
 \caption{Effect of loading direction $\gamma$ on mechanical response of the perforated sheets.
Evolution of the (a) normalized stress $\sigma_x/\bar{E}$, (b) the in-plane macroscopic Poisson's ratio $\nu_{yx}$ and (c) the opening angle of cuts $2\theta_{o_1}$ and $2\theta_{o_2}$ as a function of the applied strain $\varepsilon_x$ for different values of $\gamma$.
Note that $\nu_{yx}$ is negative only for $\varepsilon_x<\varepsilon_c$ (as at this stage the deformation of the structure is purely planar and identical to that of a network of rotating squares) and that it increases sharply and reaches positive values once the instability is triggered.
(d) Numerical snapshots of 3D patterns obtained at $\varepsilon_x=0.125$ for different values of $\gamma$.
The contours shows the normalized out-of-plane displacements.}
\label{Fig4}
\end{figure*}

Third, for large enough values of the applied strain $\varepsilon_x$, the stress $\sigma_x$ rises sharply again.
This regime starts when the square domains align [Fig.~\ref{Fig2}(b) at $\varepsilon_x=0.24$] and the deformation mechanism of the hinges switches from bending- to stretching-dominated.
At this stage, localized zone of intense strain (of plastic nature) develop in the hinges and result in the formation of permanent folds.
Although we start with a flat elastic sheet with an embedded array of cuts (i.e.~a perforated sheet), by largely stretching it we form a system that comprises a periodic distribution of both cuts and folds (i.e.~a kirigami sheet).
In particular, we note that our kirigami sheets possess several deformation characteristics of the Miura-ori~\cite{Schenk2013, Wei2013} and zigzag-base folded kirigami~\cite{Eidini2015,Eidini2016} (see Movie~3 in Supplemental Material~\cite{SI}), as ($i$) they are flat-foldable [Fig.~\ref{Fig3}(a)]; ($ii$) they form a saddle shape with a negative Gaussian curvature upon non-planar bending [Fig.~\ref{Fig3}(b)]; and ($iii$) they can be twisted under anti-symmetric out-of-plane deformation, [Fig.~\ref{Fig3}(c)].
However, in contrast to the Miura-ori, misaligned Miura-ori and zigzag-base folded kirigami, the macroscopic Poisson's ratio of our kirigami sheets is positive (see Movie~4 in Supplemental Material~\cite{SI}).
This is the result of the fact that not all the faces are rigid.
As such, the applied tensile deformation not only results in the rotation of the faces about the connecting ridges, but also in the deformation of those defined by the cuts, allowing lateral contraction of the structure.
It is also noteworthy that, differently from the misaligned Miura-ori that can only be folded to a plane, the additional degree of freedom provided by the open cuts allow the Miura kirigami to be laterally flat-foldable [Movie~4].
Finally, we note that our Miura kirigami structures have higher bending rigidity than the corresponding flat perforated sheet (see Fig.~\ref{Fig3}(d) and Movie~3 in Supplemental Material~\cite{SI}).


Having determined that instabilities in thin sheets with an embedded array of mutually perpendicular cuts can be harnessed to form complex 3D patterns,
we further explore the design space using Finite Element (FE) analyses (See Supplemental Material: {\it FE Simulations}~\cite{SI}).
We start by numerically investigating the response of finite size samples stretched along the square diagonals (i.e.~$\gamma=45^\circ$) and find excellent agreement with the experimental results (Fig.~S5 and Movie~2 in Supplemental Material~\cite{SI}). This validates the numerical analyses and indicates that they can be effectively used to explore the response of the system. First, we use the simulations to understand how plastic deformation evolves. By monitoring the distribution of the von Mises stress within the sheets, we find that that plastic deformation initiates at the tip of hinges well after the buckling onset  [see Figs.~S6~and~S9] and then gradually expand to fully cover the hinges  when the sample is fully stretched and the deformation mechanism changes from bending-dominated to stretching-dominated.
Second, we numerically explore the effect of different loading conditions and find that uniaxial tension is the ideal one to trigger the formation of well-organized out-of-plane patterns in our perforated sheets [see Fig. S7]. Third, we investigate the effect of the loading direction by simulating the response of periodic unit cells.
In Fig.~\ref{Fig4}(a) we report the stress-strain responses obtained numerically for perforated sheets characterized by $t/\delta=0.127$ and $\delta/l=0.04$ loaded uniaxially for $\gamma=0^\circ$, $15^\circ, 30^\circ$ and $45^\circ$.
Our results indicate that the mechanical response of the perforated sheets under uniaxial tension is  minimally affected by the loading direction. In fact, the evolution of both stress [Fig.~\ref{Fig4}(a)] and macroscopic in-plane Poisson's ratio [Fig.~\ref{Fig4}(b)] are similar for different values of $\gamma$.
By contrast, we find that  the morphology of the 3D patterns induced by the instability is significantly affected by $\gamma$ [Fig.~\ref{Fig4}(c)-(d)].
As the loading directions varies from $\gamma=45^\circ$ to $\gamma=0^\circ$, the symmetry in opening angle of the two sets of perpendicular cuts breaks.
While for $\gamma=45^\circ$  all cuts open equally (i.e. $\theta_{o1}=\theta_{o2}$), as we reduce $\gamma$, one set becomes wider (i.e. $\theta_{o1}$ monotonically increases) and the other progressively narrower (i.e. $\theta_{o2}$ monotonically decreases)[Fig.~\ref{Fig4}(c)].
In the limit case of $\gamma=0^\circ$ one set of cuts remains almost closed and a 3D cubic pattern emerge after buckling [Fig.~\ref{Fig1}(d), Movie~5].
Furthermore, permanent folds with direction controlled by $\gamma$ can be introduced by largely stretching the perforated sheets.
As such, by controlling the loading direction a variety of kirigami sheets can be formed [Movie~6].
While all of them are laterally flat-foldable, we find that by increasing $\gamma$ from $0^\circ$ to $45^\circ$ the resulting kirigami sheets have higher bending rigidity and their Gaussian curvature varies from zero (for $\gamma=0^\circ$) to large negative values (for $\gamma=45^\circ$).
Furthermore, by increasing $\gamma$, the resulting kirigami sheets become more compliant under torsion (Movie~6 in Supplemental Material~\cite{SI}).

In summary, our combined experimental, analytical and numerical study indicates that buckling in thin sheets perforated with a square array of cuts and subjected to uniaxial tension can be exploited to form 3D patterns and even create periodic arrangements of permanent folds.
While buckling phenomena in cracked thin plates subjected to tension have traditionally been regarded as a route toward failure~\cite{Zielsdorff1972}, we show that they can also be exploited to transform flat perforated sheets to kirigami surfaces.
Our buckling-induced  strategy not only provides a simple route for manufacturing  kirigami sheets, but can also be combined with optimization techniques to design perforated patterns  capable of generating desired  complex 3D surfaces  under external loading~\cite{Sussman2015,Dudte2016,Konakovic2016}.
Finally, since the response of our perforated sheets is essentially scale-free, the proposed {\em pop-up} strategy can be used to fabricate kirigami sheets over a wide range of scales, from transformable meter-scale architectures to tunable nano-scale surfaces~\cite{Cavallo2014,Wu2016}.

\section*{Acknowledgements}
K.B. acknowledges support from the National Science Foundation under grant number DMR-1420570 and CMMI-1149456.
A.R. also acknowledges the financial support provided by Swiss National Science Foundation (SNSF) under grant number 164648. The authors thank Bolei Deng for fruitful discussions, Yuerou Zhang for assistance in laser cutting and Matheus Fernandes for proofreading the manuscript.


\newpage
\setcounter{figure}{0}
\setcounter{equation}{0}

\makeatletter
\renewcommand{\thefigure}{S\@arabic\c@figure}
\renewcommand{\thetable}{S\@arabic\c@table}
\renewcommand{\theequation}{S\@arabic\c@equation}
\makeatother

\section{SUPPLEMENTAL MATERIALS}

\subsection{Analytical Exploration}

To get a deeper understanding of the mechanical response of the considered patterned sheets, we analytically investigate their behavior. We first study the initial in-plane linear elastic response of the system and then characterize the onset of instability resulting in the formation of 3D patterns. In all our calculations we assume that all deformation is localized at the hinges and that the square domains are rigid.

{\bf Initial in-plane linear elastic response}.
The stress-strain curves shown in Fig. 2a of the main text show that the response of all samples is initially linear.
Here, we derive an analytical relation for the effective Young's modulus of the perforated sheets, $\bar E$, in terms of the geometrical parameters  $l$ and $\delta$, and the Young's modulus of the sheet $E$.

\begin{figure}[b]
\centering
\linespread{1}
\includegraphics [width=\columnwidth]{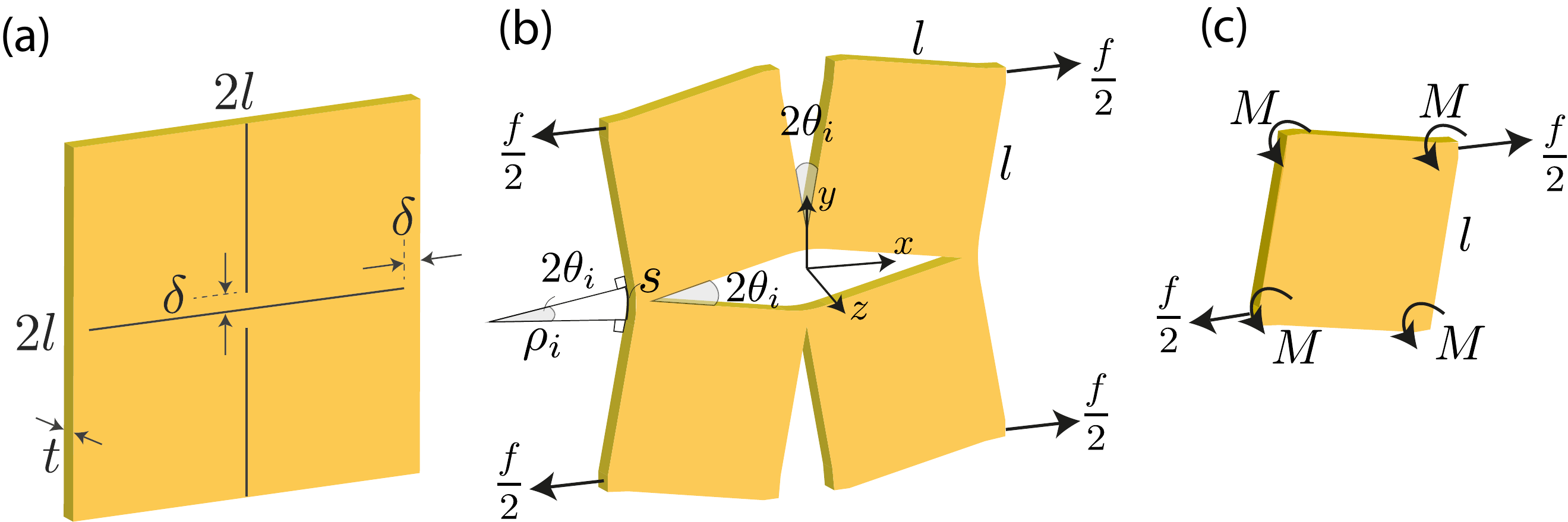}
 \caption{(a) Schematic of the system in the undeformed configuration: an elastic sheet
of thickness $t$ perforated with a square array of mutually
orthogonal cuts. (b) Schematic of the system in the planarly deformed configuration (c) Free body diagram of a square domain. }
\label{FigS1}
\end{figure}

We focus on  a unit cell comprising four square domains [Fig.~\ref{FigS1}(a)] and deform it uniaxially along one set of cuts (i.e. $\gamma=0^\circ$) by applying a macroscopic  stress $\sigma_x=f/(2 l t)$, where $f$ is the force applied to the hinges  on the vertical boundaries  and $2lt$ denotes its cross sectional area  [Fig.~\ref{FigS1}(b)]. It is important to note that using standard axis transformation techniques it has been shown that the planar response of such perforated sheets is not affected by the loading direction $\gamma$~\cite{Vasiliev2002, Grima2005}. Therefore, although here for the sake of simplicity we consider $\gamma=0^\circ$,  we expect $\bar E$ to be identical for any loading direction (i.e. for any value of $\gamma$).

The applied uniaxial stress generates identical bending moments $M$ at all  hinges, which in turn induce the rotation of all square domains by an angle $\theta_i$ and the opening of the all cuts by an angle $2\theta_i$ [Fig.~\ref{FigS1}(b)]. Focusing on a single square domain [Fig.~\ref{FigS1}(c)], it is easy to see that:
\begin{equation}
\label{Eq_S1}
M=\frac{1}{4}\times\frac{fl}{2}=\frac{1}{4}\sigma_x l^2 t.
\end{equation}
Moreover, since $M$ must be balanced by the couple induced by the internal loads, we have:
\begin{equation}
\label{Eq_S2}
M=\frac{EI_i}{\rho_i},
\end{equation}
 where  $I_i=\delta^3 t/12$ is the second moment of area of each hinge about the $z$-axis and $\rho_i$ denotes the curvature of each bent hinge.
Assuming that the length of the bent region of the  hinge  is approximately equal to the hinge width $\delta$, we obtain:
\begin{equation}
\rho_i\simeq \frac{\delta}{2\theta_i}
\label{Eq_S3}
\end{equation}

Substitution of Eqs. (\ref{Eq_S1}) and (\ref{Eq_S3}) into Eq. (\ref{Eq_S2}) yields:
\begin{equation}\label
{Eq_S4}
\sigma_x=\frac{2}{3} E \left (\frac{\delta}{l} \right)^2 \theta_i
\end{equation}

Moreover, since the strain in the loading direction is  given by:
\begin{equation}
\label{Eq_S5}
\varepsilon_x=\cos\theta_i+\sin\theta_i-1,
\end{equation}
in the small deformation regime (i.e. $\theta_i \rightarrow 0$) we have:
\begin{equation}
\label{Eq_S6}
\varepsilon_x\simeq\theta_i,
\end{equation}
so that
\begin{equation}
\label{Eq_S7}
\sigma_x=\frac{2}{3} E \left (\frac{\delta}{l} \right)^2 \varepsilon_x
\end{equation}
It follows that the effective Young's modulus of perforated sheet, $\bar E$, is:
\begin{equation}
\label{Eq_S8}
\centering
\bar{E}=\frac{\sigma_x}{\varepsilon_x}=\frac{2}{3}E \left(\frac{\delta}{l} \right)^2.
\end{equation}

Finally, we note that considering each hinge as a beam of thickness $\delta$ and width $t$, its  strain energy density under in-plane deformation can be calculated as:
\begin{equation}
\label{Eq_S9}
\mathcal{U}_{hinge}=\frac{1}{2V}\int_0^\delta \frac{E I_i}{ \rho_i^2} ds=\frac{1}{16}\bar{E}\varepsilon_x^2
\end{equation}
where $V=4l^2 t$ is the volume of the unit cell.
The in-plane strain energy density of a unit cell comprising eight hinges is then given by:
\begin{equation}
\centering
\mathcal{U}_i=8\times\mathcal{U}_{hinge}=\frac{1}{2}\bar{E}\varepsilon_x^2
\label{Eq_S10}
\end{equation}

\begin{figure}[ht]
\centering
\linespread{1}
\includegraphics [width=\columnwidth]{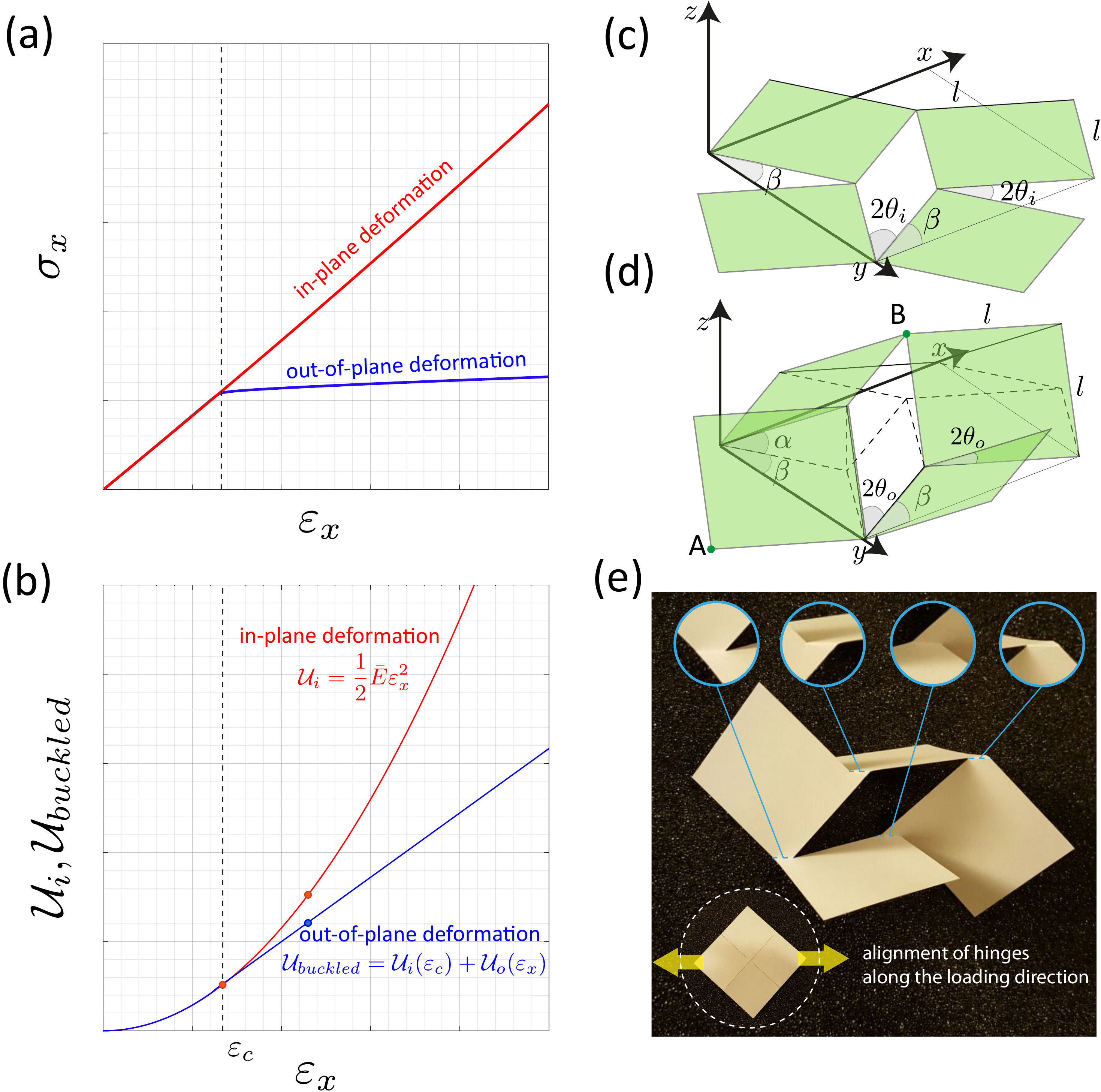}
 \caption{(a) Schematic of the stress-strain relationship for a perforated sheet that deform in-plane (red line) and out-of-plane (red line). (b) Schematic of the evolution of the strain energy versu the applied strain  for a perforated sheet that deform in-plane (red line) and out-of-plane (red line).
(c) Schematic of a  unit cell deforming in-plane. (d) Schematic of a  unit cell deforming out-of-plane.
(e) Paper model of the unit cell deformed out-of-plane.  The model clearly shows that the hinges are aligned along the loading direction. }
\label{FigS2}
\end{figure}

{\bf Transition from in-plane to out-of-plane response.}
The stress-strain curves reported in Fig.~1a of the main text also show that the thin samples (i.e.~$t/\delta\ll1$) are characterized by
a sudden departure from linearity to a plateau stress.
This sudden departure from linearity is the result of out-of-plane buckling of the ligaments and occurs when the out-of-plane deformation of the hinges becomes energetically less costly than their in-plane deformation.
While the  strain energy density of a  perforated sheet that has deformed in-plane is given by Eq.~(\ref{Eq_S10}),
the strain energy density of a  perforated sheet that has deformed out-of plane can be expressed as:
\begin{equation}
\centering
\mathcal{U}_{buckled}(\varepsilon_x)=
             \mathcal{U}_i(\varepsilon_c)+\mathcal{U}_o(\varepsilon_x)
\label{Eq_S11}
\end{equation}
where $\varepsilon_c$ is the critical strain at which buckling occurs and $\mathcal{U}_o$ is energy contribution due to the  out-of-plane deformation.

Since  $\sigma_x=d\mathcal{U}_{buckled}/d\varepsilon_x$ and the stress  immediately after instability is almost constant [Fig.~\ref{FigS2}(b)], it follows that $\mathcal{U}_o(\varepsilon_x)$ is approximately linear in $\varepsilon_x$  [Fig.~\ref{FigS2}(a)] and can be identified as the tangent of $\mathcal{U}_i(\varepsilon_x)$ at $\varepsilon_c$:

\begin{equation}
\centering
\mathcal{U}_o(\varepsilon_x)=\bar{E}\varepsilon_c(\varepsilon_x-\varepsilon_c)
\label{Eq_S12}
\end{equation}

Moreover, as all hinges bend out-of-plane, $\mathcal{U}_o(\varepsilon_x)$ can also be expressed as:

\begin{widetext}
\begin{equation}
\mathcal{U}_o(\varepsilon_x)=4\times\frac{1}{2V}\int_0^\delta  \frac{E I_o}{\rho_{o_1}^2} ds+4\times\frac{1}{2V}\int_0^\delta  \frac{E I_o}{\rho_{o_2}^2} ds= 
 \frac{1}{6} E\left(\frac{t}{l} \right)^2(\theta_{o_1}^2+\theta_{o_2}^2)
\label{Eq_S13}
\end{equation}
\end{widetext}
where $I_o=\delta t^3/12$ is the second moment of area with respect to $x$-axis ($y$-axis) for hinges along $y$-axis ($x$-axis), $ \rho_{o_j}=\delta/(2\theta_{o_j})$ is the out-of-plane curvature of the deformed hinges and $2\theta_{o_j}$ is the opening angle of an hinge after  out-of-plane buckling. Note that, in general, after buckling the opening angles of the hinges within the perforate sheet can take two values,  $2\theta_{o_1}$ and $2\theta_{o_2}$ (i.e. not all hinges open equally after buckling).

While Eq. (\ref{Eq_S13}) is valid for any loading direction, for $\gamma=45^o$ (i.e. loading along the square diagonals) $\theta_{o_1}=\theta_{o_2}=\theta_{o}$, so that Eq. (\ref{Eq_S13}) simplifies to:
\begin{equation}
\centering
\mathcal{U}_o(\varepsilon_x)=8\times\frac{1}{2V}\int_0^\delta  \frac{E I_o}{\rho_o^2} ds=\frac{1}{3}E\left(\frac{t}{l} \right)^2\theta_o^2.
\label{Eq_S14}
\end{equation}
The critical strain $\varepsilon_c$ can then be determined by equation Eqs. (\ref{Eq_S12}) and (\ref{Eq_S14}) after having expressed $\theta_o$ as a function of $\varepsilon_x$. To this end,
we first note that for  $\gamma=45^o$  [Fig.~\ref{FigS2}(d)]
\begin{equation}
\centering
\cos 2\theta_{o} = \cos \alpha \sin 2\beta
\label{Eq_S15}
\end{equation}
where $\alpha\in [0,\pi/2]$ determines the orientation of the square domains with respect to $xy$-plane and
\begin{equation}
\centering
\beta=\arccos\frac{1+\varepsilon_x}{\sqrt2}.
\label{Eq_S17}
\end{equation}
is the angle  between the edge of the square domain  placed on the $xy$-plane and the $x$-axis.

Moreover, since the experiments indicate that for $\gamma=45^\circ$ the hinges are approximately aligned along the loading direction [see paper illustration in Fig.~\ref{FigS2}(e)], the distance between  points A and B shown in Fig.~\ref{FigS2}(d) remains constant (i.e. $\rvert\overline{AB}\rvert=\sqrt{5}l$) and
\begin{equation}
\cos \alpha=\tan\beta.
\label{Eq_S17a}
\end{equation}

While Eq. (\ref{Eq_S17a}) is exact well into the postbuckling regime, it does not correctly capture the response of the system at the onset of instability, as it predicts  $\alpha\neq0$ for $\varepsilon_c$ (i.e. it predicts that the square are already rotated out of  the $xy$-plane when the instability is triggered). To correct for this, we modify Eq. (\ref{Eq_S17a}) as
\begin{equation}
\cos \alpha\simeq\tan(\beta+\theta_c), \;\; \textrm{for }  \varepsilon_x\geq\varepsilon_c.
\label{Eq_S17b}
\end{equation}
where $\theta_c=\pi/4-\beta_c$  denotes the opening angle associated to $\varepsilon_c$. Note that, according to Eq. (\ref{Eq_S17b}), $\alpha=0$ at $\varepsilon_c$
Substitution of Eqs. (\ref{Eq_S17}) and (\ref{Eq_S17b}) into Eq. (\ref{Eq_S15}) yields:
\begin{widetext}
\begin{equation}
\theta_o(\varepsilon_x) =
\frac{1}{2} \arccos
 \left[\sin\left(2\arccos \frac{1+\varepsilon_x}{\sqrt2}\right)
\tan\left(\frac{\pi}{4}+\arccos \frac{1+\varepsilon_x}{\sqrt2}-\arccos \frac{1+\varepsilon_c}{\sqrt2}\right)\right ]
\label{Eq_S18}
\end{equation}
\end{widetext}
Although Eq. (\ref{Eq_S18}) provides a highly non-linear relation between $\theta_o$ and $\varepsilon_x$, close to the instability point $\theta_o^2(\varepsilon_x)$ can be approximated as
\begin{equation}
\centering
\theta_o^2(\varepsilon_x) \simeq\varepsilon_x-\varepsilon_c
\label{Eq_S18a}
\end{equation}
and can be then inserted into Eq. (\ref{Eq_S11}) to express $U_o$ in terms of $\varepsilon_x$.
Finally, the critical strain $\varepsilon_c$ can be determined by substituting Eq. (\ref{Eq_S18a}) into Eq.~(\ref{Eq_S14}) and then equating it to Eq.~(\ref{Eq_S12}):

\begin{equation}
\centering
 \varepsilon_c \simeq\frac{1}{2} \left ( \frac{t}{\delta} \right )^2
\label{Eq_S21}
\end{equation}

This relation shows that the critical strain scales quadratically with $t/\delta$ and despite several simplifications made, it compares very well with both experimental [Fig.~2(c)] and numerical [Fig.~\ref{FigS8}] results.

\subsection{Experiments}
\begin{figure}[b]
\centering
\linespread{1}
\includegraphics [width=0.6\columnwidth]{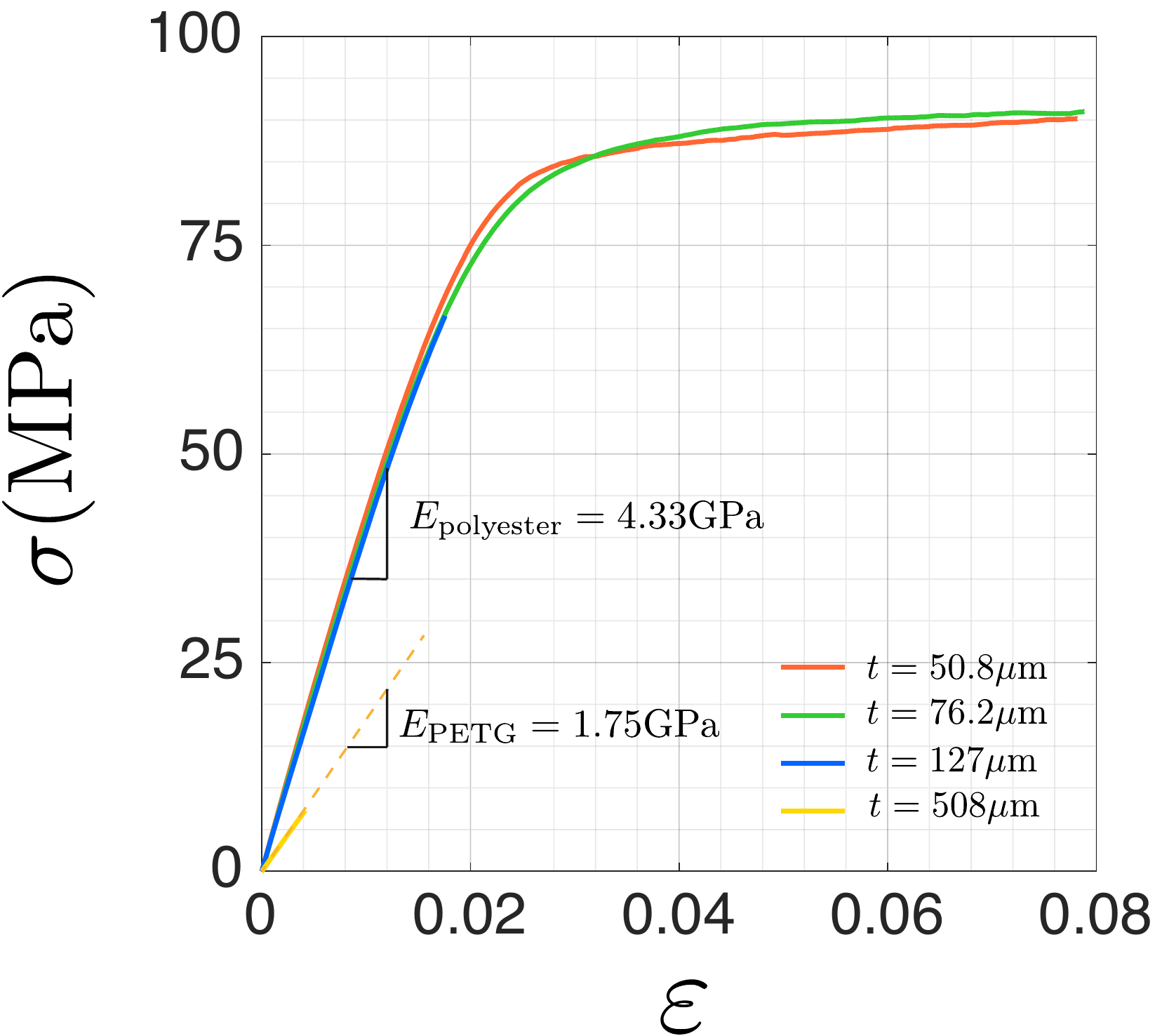}
 \caption{ Stress-strain curves of the plastic sheets obtained by stretching them uniaxially. The sheets with $t=50.8$, $76.2$ and $127\mu$m are made of polyester, while the one with $t=508\mu$m is made of PETG.  }
\label{FigS3}
\end{figure}

{\bf Fabrication}.
Specimens are fabricated by laser cutting an array of $3\times8$ mutually perpendicular cuts into plastic sheets (Artus Corporation, NJ). Total number of 10 samples are fabricated with a combination of different sheet thickness ($t=50.8\mu$m, $76.2\mu$m, $127 \mu$m and $508\mu$m) and three normalized hinge widths ($\delta/l=0.06,0.1,0.14$). Note that in all our samples $l=10$mm.
The material properties of the  plastic sheets  used in this study are characterized by performing uniaxial tensile tests (ASTM D882) with a uniaxial testing machine (Instron 5566) equipped with a 100N load cell.
Strips with a width of 10 mm and gauge length of 120 mm are fully clamped at both ends using a pneumatic gripper and stretched with a displacement rate of 0.1 mm/s up to $\varepsilon=0.8$ [Fig.~\ref{FigS3}]. The stress-strain curves reported in Fig.~\ref{FigS3} indicate that the polyester sheets with thickness $t=50.8\mu$m (red line), $76.2\mu$m (green line), $127 \mu$m (blue line) are characterized by a Young's modulus  $E=4.33$ GPa. Moreover, their 0.2\% offset yield strength is measured as $\sigma_y=66.4$ MPa and their plastic strain $\varepsilon_p$ versus $\sigma$ is reported in the Table~\ref{TableS1} up to fully plastic region. Differently, for the thick PETG sheet ($t=508\mu$m, yellow line)
we measure a Young's modulus  $E=1.75$ GPa. Finally, we note that for all the plastic sheets a typical Poisson's ratio $\nu=0.4$ is assumed.

\begin{table}[ht]
\begin{center}
  \begin{tabular}{ c||c|c|c|c|c|c|c}
    \hline
    $\varepsilon_p$     & 0      & 0.002 & 0.004 & 0.006 & 0.008 & 0.018 & 0.028 \\ \hline
    $\sigma$ (MPa) & 66.4 & 75.78 & 80.52 & 82.75 & 84.45 & 88.09 & 89.06 \\
    \hline
  \end{tabular}
\end{center}
\caption{Plastic strain ($\varepsilon_p$) versus stress ($\sigma$) for the three polyester sheets.}
\label{TableS1}
\end{table}

\begin{figure}[t]
\centering
\linespread{1}
\includegraphics [width=0.75\columnwidth]{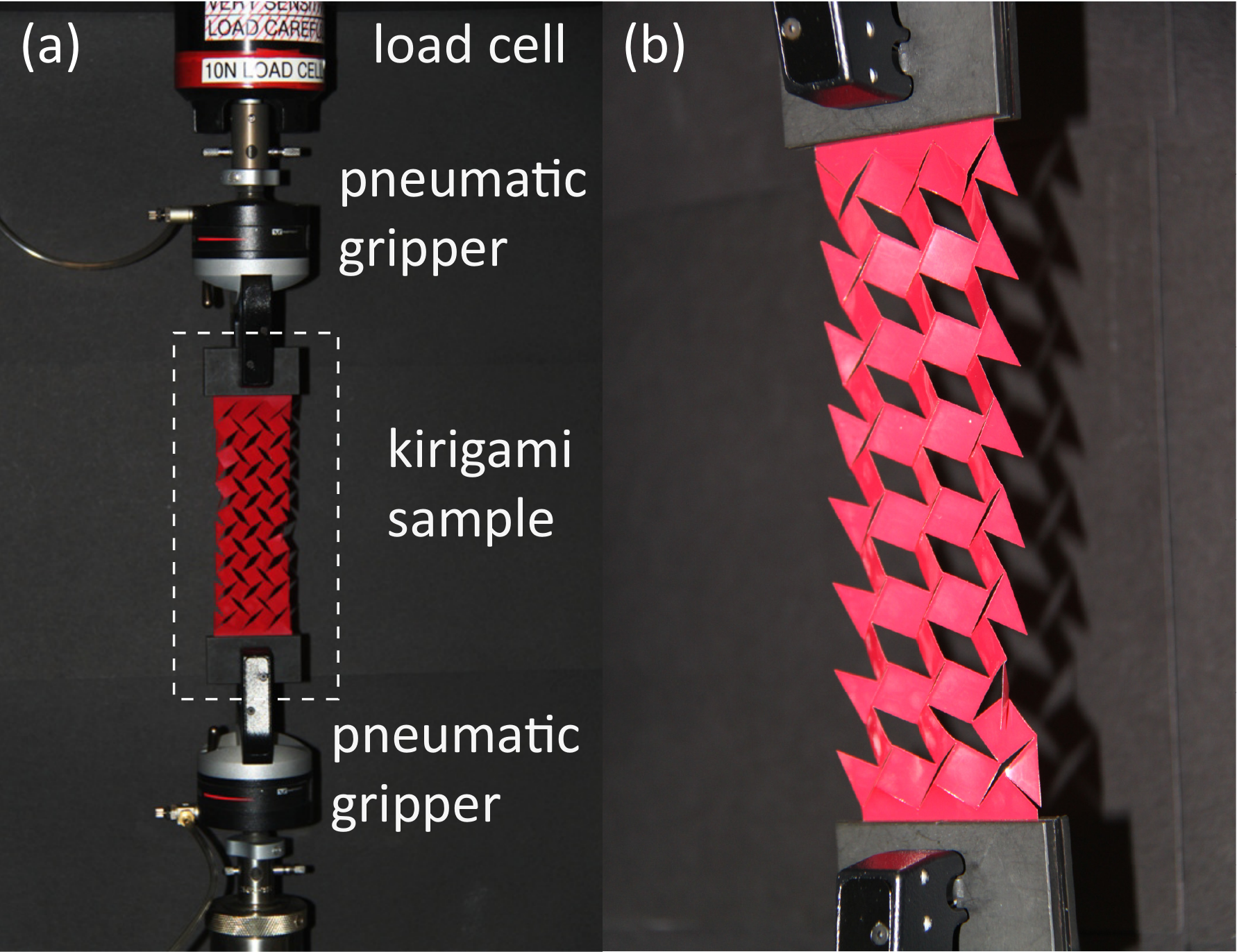}
 \caption{ (a) Experiment setup and (b) closeup view of the perforated sheet.  }
\label{FigS4}
\end{figure}

{\bf Testing.}
The quasi-static uniaxial tensile response of the perforated sheets comprised of $3\times8$ unit cells was probed by using a uniaxial testing machine (Instron 5566) equipped with a 10N load cell. All test were conducted under displacement rate of $\dot{u}_x=0.1$ mm/s [Fig.~\ref{FigS4}].

\subsection{Finite Element simulations}
In this Section, we provide details about the Finite Element (FE) simulations conducted for this study using the commercial package Abaqus\textbackslash Standard 6.12 (Dassault Syst\`emes).
In all simulations, the models are discretized with 3D shell elements (S4R) and the cuts in flat sheets are modeled as seam cracks with duplicate overlapping nodes along the cuts.

{\bf Finite size simulations}.
To validate the FE simulations, we first performed finite size simulations on perforated sheets similar to those used in experiments comprising an array of $3\times8$ cells.
The lower boundary of the sample is fixed and a vertical displacement is applied to the upper boundary while the lateral boundaries are assumed to be traction free.
The material behavior of the plastic sheet is captured using an elasto-plastic model (material models *ELASTIC and *PLASTIC in Abaqus) with the experimentally characterized properties (see Fig.~\ref{FigS3} and Table~\ref{TableS1}).
The response of the sheets is then simulated conducting dynamic implicit simulations (*DYNAMIC module in Abaqus).
To facilitate convergence, we introduce some artificial, numerical damping (by setting the parameters $\alpha=-0.41421$, $\beta=0.5$ and $\gamma=0.91421$ in the Hilber-Hughes-Taylor time integration algorithm).
Moreover, quasi-static conditions are ensured by monitoring the kinetic energy and finally, to trigger the instability an imperfection is introduced by applying two opposing small bias forces normal to the sheet plane at two ends of each cut during the initial phase of each simulation.

\begin{figure}[t]
\centering
\linespread{1}
\includegraphics [width=\columnwidth]{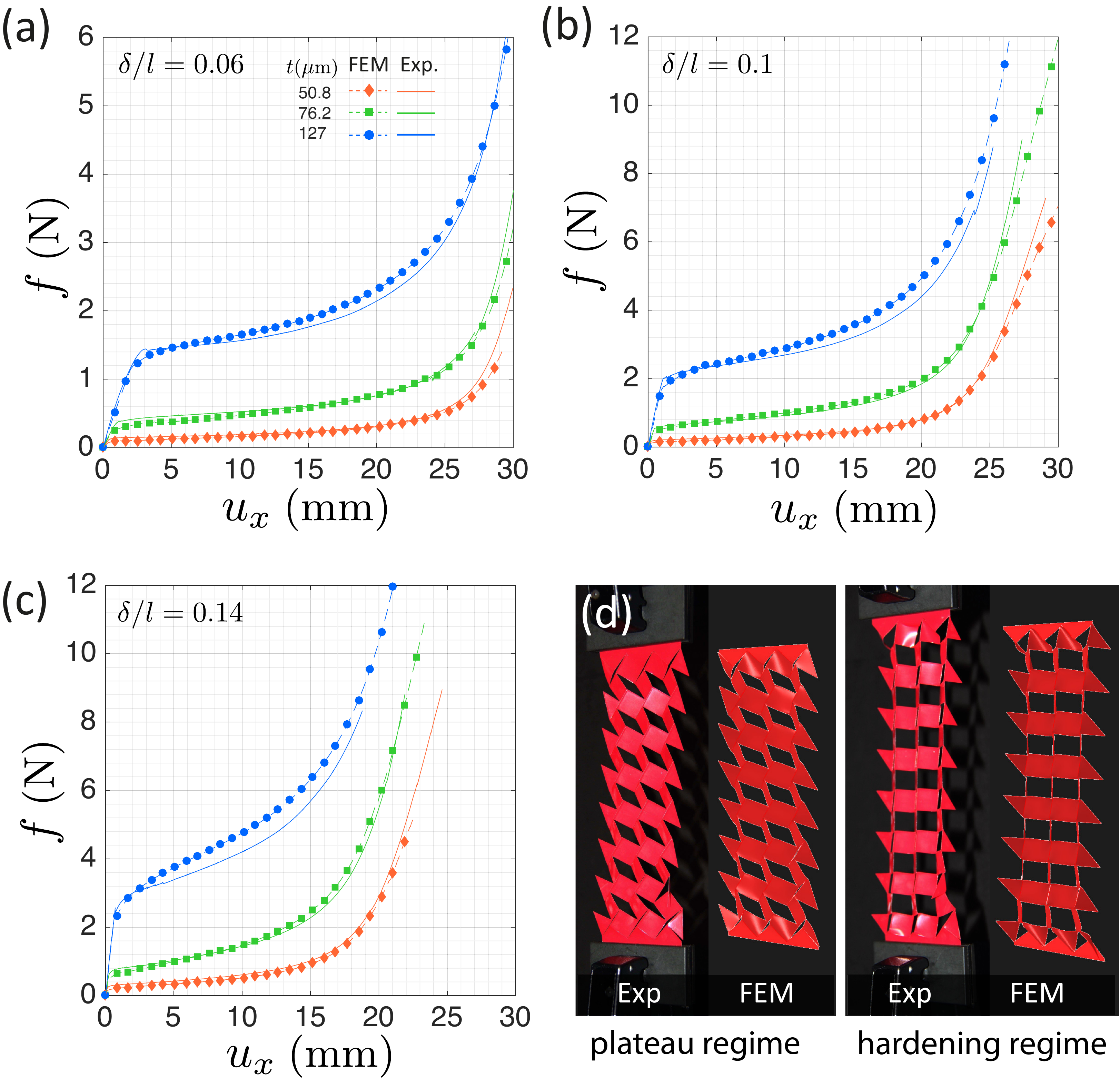}
 \caption{ Comparison between experimental and numerical force displacement curves for (a) $\delta/l=0.06$, (b) $\delta/l=0.1$ and (c) $\delta/l=0.14$ and different thickness $t$ where $l=10$mm.
 (d) Snapshots of experimental and numerical results at $\varepsilon_x=0.125$. }
\label{FigS5}
\end{figure}

\begin{figure}[b]
\centering
\linespread{1}
\includegraphics [width=\columnwidth]{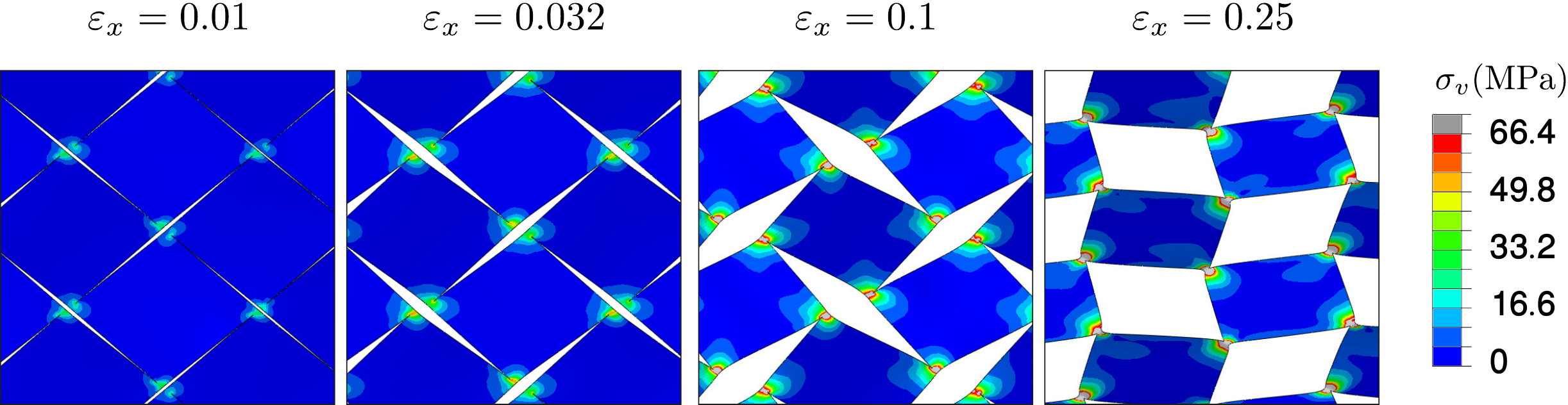}
 \caption{Von Mises stress distribution $\sigma_v$ at the hinges located in the  middle of a finite size sample characterized by $t/\delta=0.06$ ($l=10$mm and $t=50.8\mu$m) for $\gamma=45^\circ$. Snapshots  at different strain levels are shown. Note the von Mises stress  reaches the yield strength of the material ($\sigma_y=66.4$ MPa) when $\varepsilon_x=0.032$. Gray contours show the regions where $\sigma_v$ is larger than the yield strength $\sigma_y$ (i.e. plastic zones).}
\label{FigS6}
\end{figure}

First, we numerically investigate the response of finite size samples stretched along the square diagonals (i.e.~$\gamma=45^\circ$) and find excellent agreement with the experimental results (Fig.~\ref{FigS5} and Movie~2). This validates the numerical analyses and indicates that they can be effectively used to explore the response of the system. Moreover, the simulations provide additional insights, since they allow us to easily monitor the stress distribution within the sheets and, therefore, to understand how plastic deformation evolves.
In Fig.~\ref{FigS6} we show a close-up view of the distribution of von Mises stress, $\sigma_v$, at the hinges located in the middle of a finite size sample characterized by $t/\delta=0.06$ ($l=10$mm and $t=50.8\mu$m) for $\gamma=45^\circ$. Since for the material considered in this study plastic deformation develops when $\sigma_v=\sigma_y=66.4$ MPa (see Fig. \ref{FigS3} and Table \ref{TableS1}), the snapshots indicate that yielding at the hinges initiate at $\varepsilon_x\simeq0.032$. Note that, although this is a very small value of strain, it is well beyond the onset of buckling (for this sample $\varepsilon_c\simeq0.0036$). We then find that the plastic zone at the hinges gradually increase with the applied strain and fully cover them  when the sample is fully stretched and the deformation mechanism of the hinges changes from bending-dominated to stretching-dominated.

Second, we use FE to explore the effect of different loading conditions.  More specifically, while in the main text we focus exclusively  on  perforated sheets subjected to uniaxial tension,  here we investigate the response of a perforated sheet characterized by $t/\delta=0.127$ and $\delta/l=0.04$ and under biaxial deformation applied at $\gamma=45^\circ$.
We perform simulations on a finite size sample comprising $3\times3$ unit cells  and consider
three load cases: ($i$) equibiaxial tension (i.e. $\varepsilon_x=\varepsilon_y>0$), ($ii$) equibiaxial compression (i.e. $\varepsilon_x=\varepsilon_y<0$) and ($iii$) biaxial tension/compression (i.e. $\varepsilon_x=-\varepsilon_y>0$). For all cases appropriate displacements in  the $x$-$y$ plane are applied to all nodes on the edges of the models, while constraining their displacements in $z$-direction (note that all rotations are left unset).

 \begin{figure}[t]
\centering
\linespread{1}
\includegraphics [width=\columnwidth]{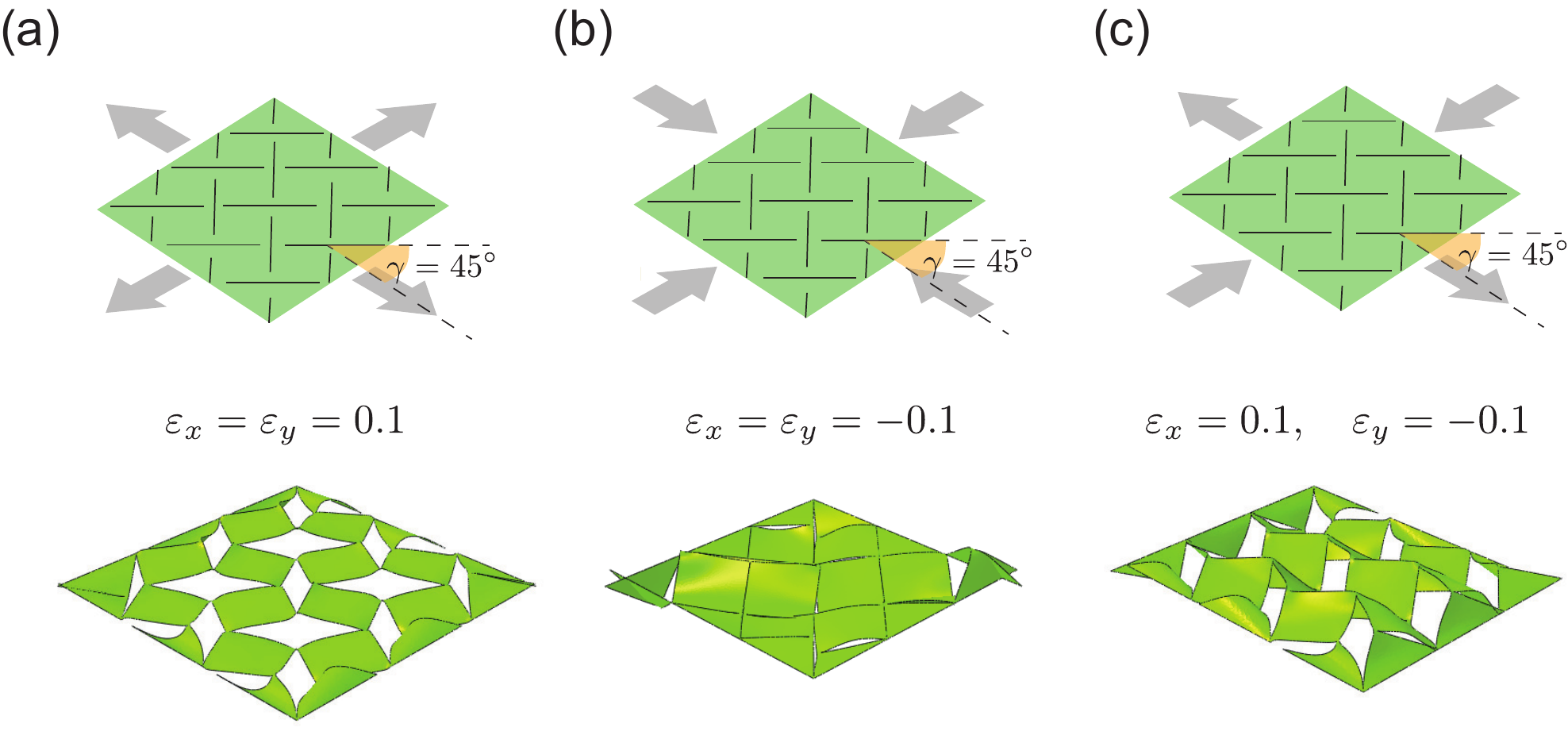}
 \caption{Numerical snapshots of perforated sheets under biaxial loading: (a) equibiaxial tension at $\varepsilon_x=\varepsilon_y=0.1$, (b) equibiaxial compression at $\varepsilon_x=\varepsilon_y=-0.1$ and (c) biaxial tension/compression at $\varepsilon_x=0.1$ and $\varepsilon_y=-0.1$.}
\label{FigS7}
\end{figure}

Our simulations indicate that under equibiaxial tension  the structure remains roughly flat and no  out-of-plane pattern emerges [see Fig.~\ref{FigS7}(a)]. This is because, as indicated by the Poisson's ratio results reported in Fig. 4(b) of the main text, the formation of the out-of-plane pattern is accompanied by lateral contraction and under equibiaxial tension such contraction is prevented by the tensile stretch applied in the transverse direction.
Moreover, as shown in Fig.~\ref{FigS7}(b), we find that under equibiaxial compression the periodic pattern of cuts does not significantly affect the response of the system. Our perforated sheet behaves similarly to a continuous thin sheet and buckles out of plane to form a dome-like shape. Finally, for the case of biaxial tension/compression our simulations show that the response  of the perforated sheets  is very similar to that observed under uniaxial tension [see Fig.~\ref{FigS7}(c)]. This is because, differently from the case of equibiaxial tension, for this loading condition the compressive stretch applied in lateral direction favors the formation of the out-of-plane pattern. However, it is important to note that in the case of  biaxial tension/compression the response of the sheets is highly affected by their size. For  sheets with larger number of unit cell aligned in the direction of the applied compressive force we find that the sheet buckles globally to form a wavy pattern. Therefore, this set of simulations indicate that uniaxial tension is the ideal loading condition to trigger the formation of well-organized out-of-plane patterns in our perforated sheets.

{\bf Unit cell simulations}.
To reduce the computational costs and make sure the response of the system is not dominated by boundary effects, we investigate the response of infinite perforated sheets under periodic boundary conditions.
Since here we are mostly interested in the response of the perforated sheet immediately after buckling (i.e. before the plastic deformation takes place), for this set of simulations we use a linear elastic material model (with $E=4.33$ GPa and $\nu=0.4$).
All  simulations consist of two steps: ($i$) we first use a linear perturbation analysis (*BUCKLE module in Abaqus) to identify the critical buckling mode; ($ii$) we then introduce a small imperfection ($\simeq0.001l$)  in the form of the critical mode into the mesh to guide the post-buckling analysis.
As for the finite size simulations, for this step we conduct dynamic implicit simulations (*DYNAMIC module in Abaqus) and to facilitate convergence, we introduce some artificial, numerical damping.

\begin{figure}[b]
\centering
\linespread{1}
\includegraphics [width=0.6\columnwidth]{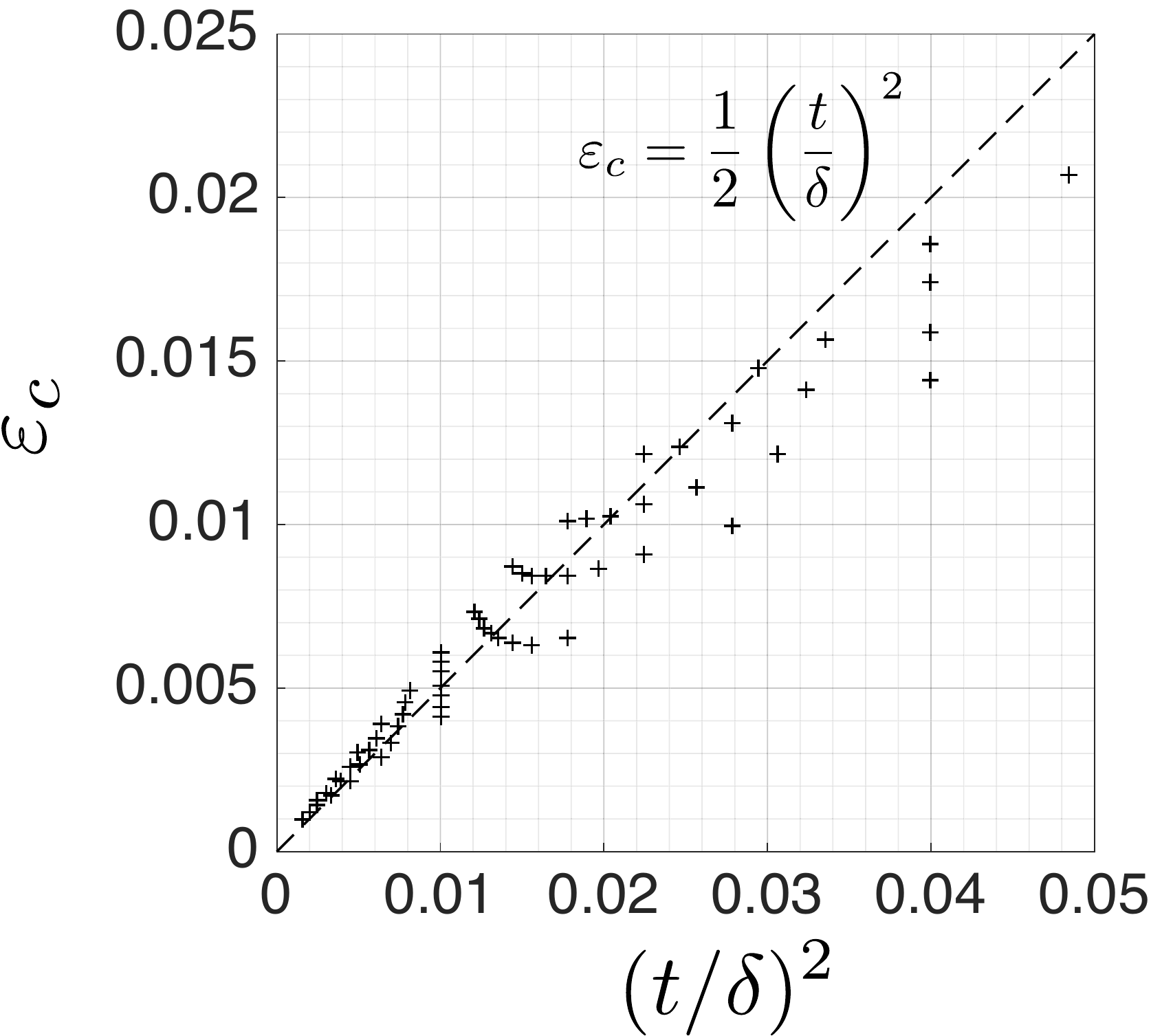}
 \caption{ Comparison of the critical strain $\varepsilon_c$ as predicted by FE simulations (markers)  and theory [dashed line - Eq.~(\ref{Eq_S21})] for $\gamma=45^\circ$.}
\label{FigS8}
\end{figure}

In Fig.~\ref{FigS8} we compare the analytical expression for the critical strain [Eq.~(\ref{Eq_S21})] with the numerical predictions of 54 unit cell simulations characterized by $\delta/l\in [0.05,0.1]$ and $t/\delta\in[0.04,0.24]$ and $\gamma=45^\circ$. We find an excellent agreement between numerical (markers) and analytical (dashed line) results.

\begin{figure}[t]
\centering
\linespread{1}
\includegraphics [width=\columnwidth]{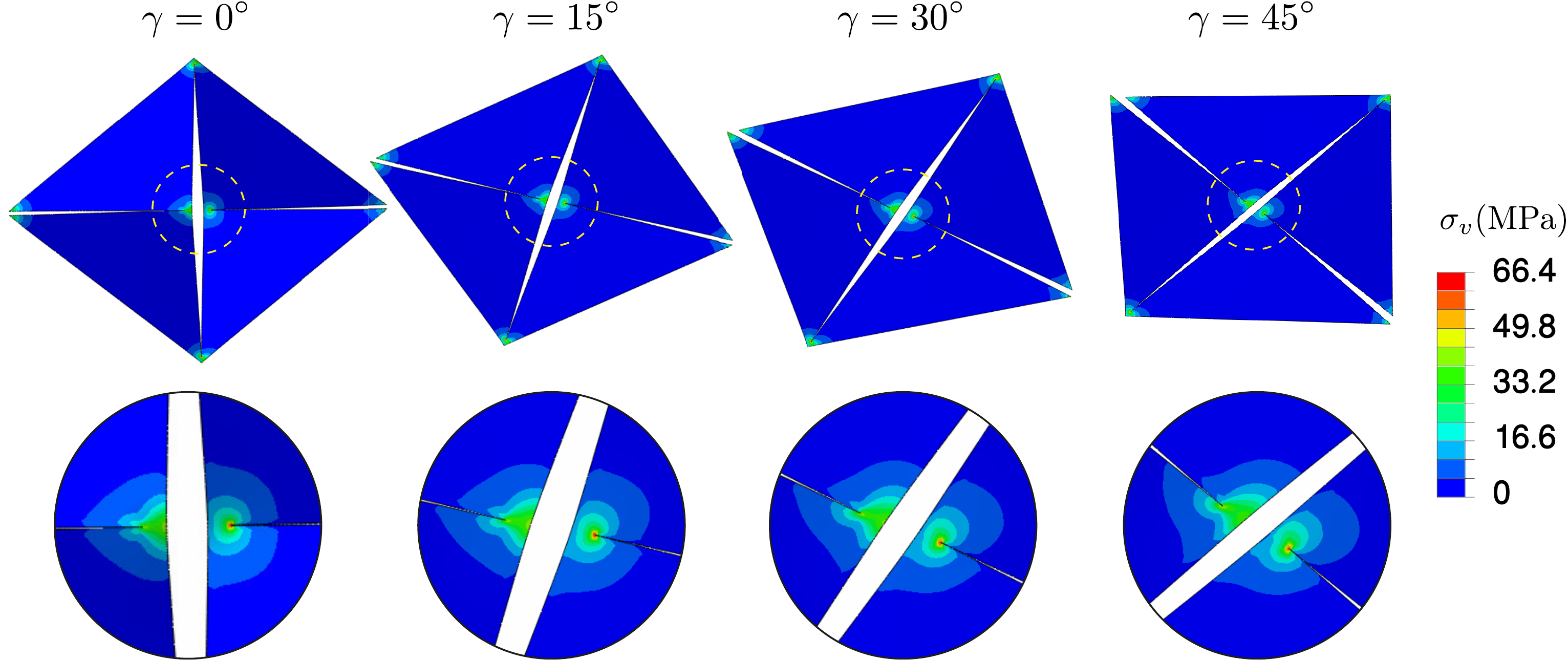}
 \caption{Numerical snapshots showing the distribution of the von Mises stress, $\sigma_v$, in periodic unit cells loaded along different orientations $\gamma$  at $\varepsilon_x=0.02$. we note that at this level of applied deformation the von Mises stress at  the tips of the cuts reach the yield strength of the sheet material ($\sigma_y=66.4$ MPa). }
\label{FigS9}
\end{figure}

Finally, in Fig.~\ref{FigS9} we show the von Mises stress distribution in a unit cell characterized by  $\delta/l=0.04$ and $t/\delta=0.127$ for different values of $\gamma$ at $\varepsilon_x=0.02$. As expected, we find that the von Mises stress, $\sigma_{v}$, is maximum at the hinges. Assuming the sheets are made of the same material used to fabricate our sample (i.e. with $E=4.33$ GPa and $\nu=0.4$), we also find that in all unit cells $\max(\sigma_{v})\sim 66$ MPa. This is the stress at which we expect plastic deformation to initiate for the considered material (see Fig.~\ref{FigS3} and Table~\ref{TableS1}). Therefore, we can conclude that  the yielding of the hinges  begins approximately at $\varepsilon_x\simeq0.02$. Since, this strain  is more than twice the critical strain (i.e. $\varepsilon_c\simeq0.008$), these simulations confirm that no plastic deformation takes place before buckling.

\subsection{Movie captions}

{\bf Movie~1} {\it Buckling-induced 3D kirigami in thin perforated sheets}.

In sufficiently thin sheets perforated with a square array of mutually orthogonal cuts mechanical instabilities are triggered under uniaxial tension and  can be exploited to create complex 3D patterns. If the sheet is loaded along the square diagonals (i.e. $\gamma=45^\circ$), a 3D pattern reminiscent of a misaligned Miura-ori  emerges, while loading along one set of the cuts (i.e. $\gamma=0^\circ$) results in a 3D cubic pattern.\\

{\bf Movie~2} {\it Uniaxial loading: Experiment vs FE simulation}.

The deformation of perforated sheets subjected to uniaxial tensile loading along the square diagonals ($\gamma=45^\circ$) can be accurately captured by FE simulations in Abaqus.\\

{\bf Movie~3} {\it Buckling-induced Miura kirigami }.

Although we start with a flat elastic sheet with an embedded array of cuts, by largely stretching it we end up with a system that comprises a periodic arrangement of both cuts and folds. As a result, after being fully stretched our sheets possess several deformation characteristics of the Miura-ori, including flat-foldability, negative Gaussian curvature under non-planar bending and twisting under anti-symmetric out-of-plane deformation.
Miura kirigami also exhibits an enhanced bending rigidity compared to a flat perforated sheet.\\

{\bf Movie~4} {\it In-plane Poisson's ratio of buckling-induced kirigami}.

In contrast to Miura-ori and misaligned Miura-ori, the in-plane Poisson’s ratio of our perforated sheets after buckling is positive.\\

{\bf Movie~5} {\it Buckling-induced cubic kirigami}.

The kirigami structure obtained by fully stretching the perforated sheet along one set of the cuts (i.e. $\gamma=0^\circ$) is ($i$) flat foldable, ($ii$) exhibits a zero Gaussian curvature under non-planar bending and ($iii$) has relatively higher torsional rigidity compared to that of the structure obtained by loading the sheet along the diagonals of the squares (i.e. $\gamma=45^\circ$).\\

{\bf Movie~6} {\it Comparing the behavior of kirigami sheets}.

Permanent folds with direction controlled by $\gamma$ can be introduced by largely stretching the perforated sheets.
As such, by controlling the loading direction a variety of kirigami sheets can be formed.
While all of them are laterally flat-foldable, we find that by increasing $\gamma$ from $0^\circ$ to $45^\circ$ the resulting kirigami sheets have higher bending rigidity and their Gaussian curvature varies from zero (for $\gamma=0^\circ$) to large negative values (for $\gamma=45^\circ$).
Furthermore, by increasing $\gamma$, the resulting kirigami sheets become more compliant under torsion.

\end{document}